# Dimensional crossover of heat conduction in amorphous Polyimide nanofibers

*Lan Dong[1,2,3†], Qing Xi[1,2,3†], Dongsheng Chen[4,5], Jie Guo[1,2,3], TsuneyoshiNakayama[1,2,3,6],*

*Yunyun Li[1,2,3], Ziqi Liang[4], Jun Zhou[1,2,3\*], Xiangfan Xu[1,2,3\*], Baowen Li[7\**

[1]Center for Phononics and Thermal Energy Science, School of Physics Science and Engineering, Tongji University, 200092 Shanghai, China

[2]China-EU Joint Center for Nanophononics, School of Physics Science and Engineering, Tongji University, 200092 Shanghai, China

[3]Shanghai Key Laboratory of Special Artificial Microstructure Materials and Technology, School of Physics Science and Engineering, Tongji University, 200092 Shanghai, China

[4]Department of Materials Science, Fudan University, 200433 Shanghai, China

[5]College of Mathematics and Physics, Shanghai University of Electric Power, 200090 Shanghai, China

[6]Hokkaido University, Sapporo 060-0826, Japan

[7]Department of Mechanical Engineering, University of Colorado, Boulder, CO 80309-0427, USA

**ABSTRACT:** The mechanism of thermal conductivity in amorphous polymers, especially polymer fibers, is unclear in comparison with that in inorganic materials. Here, we report the observation of across over of heat conduction behavior from three dimensions (3D) to quasi-one dimension (1D) in Polyimide(PI) nanofibers at a given temperature. A theoretical model based on the random walk theory has been proposed to quantitatively describe the interplay between the inter-chain hopping and the intra-chain hopping in nanofibers. This model explains well the diameter dependence of thermal conductivity and also speculates the upper limit of thermal conductivity of amorphous polymers in the quasi-1D limit.

Polymers are widely used materials due to their fascinating properties such as low mass density, chemical stability, and high malleability etc[1]. Unfortunately, the relatively low thermal conductivity of polymer, which is in the range of ~0.1 $Wm^{-1}K^{-1}$ to ~0.3 $Wm^{-1}K^{-1}$[2-4], limits its application in thermal management. The low thermal conductivity of polymer is considered to be one of the major reasons for the thermal failure in electronic devices[5,6]. Therefore, thermally conductive polymers are highly demanded for heat dissipation in microelectronic and civil applications.

Contradicting to common wisdom, polymer nanofibers hold surprisingly high thermal conductivity, some of which are even comparable to that in some metals or even silicon. Choy and his co-workers carried out the pioneer theory and experiments to demonstrate that the alignment of molecular chains could enhance the thermal conductivity along the alignment direction[7-9]. The increasing of the thermal conductivity is attributed to the increase of the degree of crystallinity in following experimental works[1,10-12]. Cai *et al*. also observed thermal conductivity enhancement in polyethylene nanowires fabricated by the improved nanoporous template wetting technique, due to the high chain orientation arising from crystallinity[13,14]. More recently, Singh *et al*. demonstrated that better molecular chain orientation could also improve the thermal conductivity when polymer fibers remain amorphous[4], which indicates that it is also of significance to study the intrinsic mechanism of thermal conductivity in amorphous polymer. All these pioneering works indicate that the thermal properties in polymers are highly related to their microscopic configurations, and thermal conductivity is limited by the molecular orientation and inter-chain scatterings[15,16].

It was also found through Molecular Dynamic(MD) simulation that the chain conformation would strongly influence thermal conductivity[17, 18]. However, very few theories have been proposed to quantitatively study the structure-dependence of thermal conductivity in amorphous polymers because of their complex intrinsic structure. Alternatively, theories for amorphous inorganic materials such as heat transfer by diffusions[19, 20], the minimum thermal conductivity model[21-23] and phonon-assisted hopping model[24-26] have been borrowed to qualitatively explain the thermal conductivity of amorphous polymers[4, 27, 28]. Compared to the unique type of hopping in inorganic amorphous materials, there are two types of hopping processes in bulk polymers, i.e. intra-chain and inter-chain hopping processes, which together with their interplay play an important role in the thermal conductance. Therefore, it is not yet totally clear the mechanism of the enhancement of thermal conductivity in polymer nanofibers and the upper limit of such enhancement when the polymer is stretched.

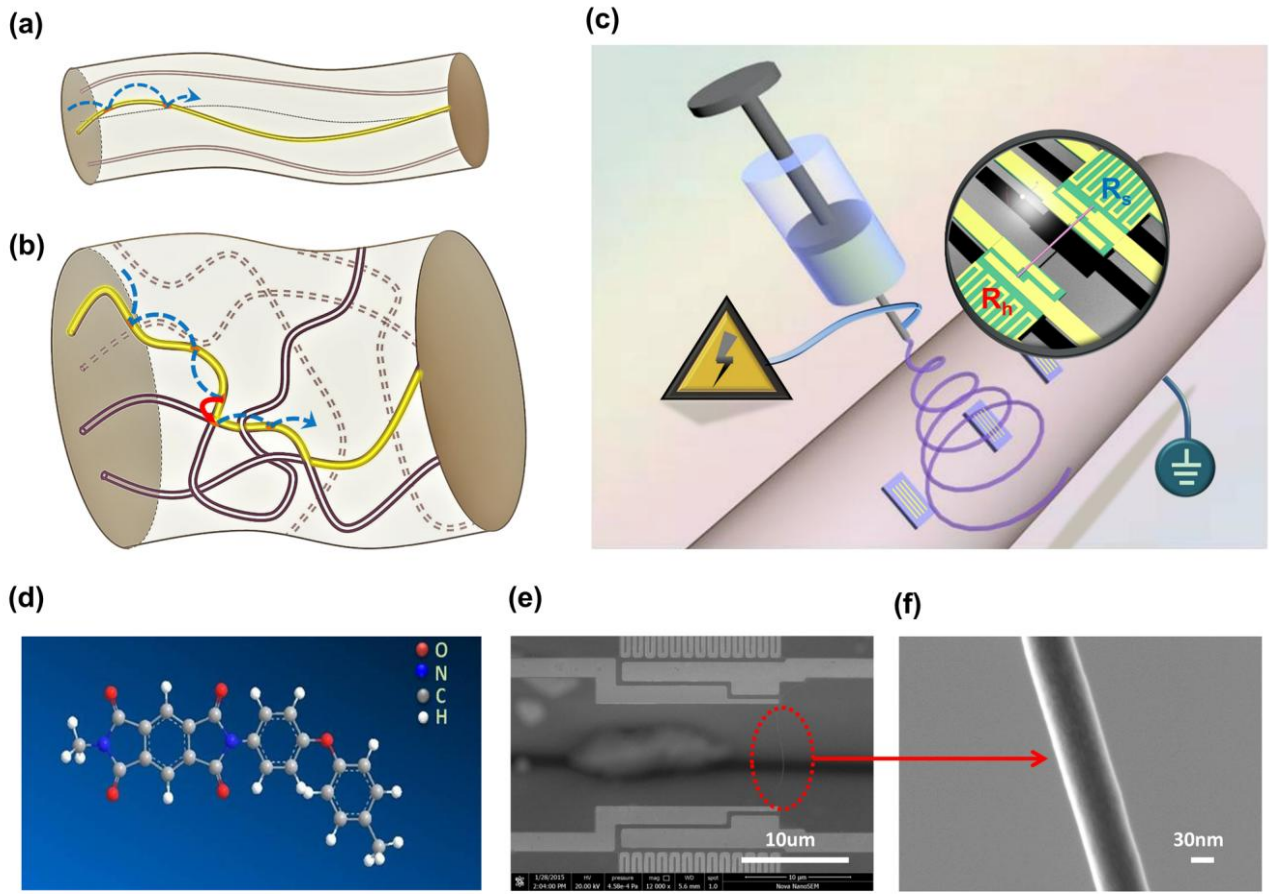

**Figure 1.** Schematic picture of the electrospinning setup and details of amorphous PI nanofibers. (a) Schematic of anisotropic quasi-one-dimensional thermal diffusion in the nanofibers with small diameters. All the molecular chains are aligned along the fiber-axis. Blue arrow denotes the hopping between neighboring localization centers within the chain and only intra-chain hopping could happen in this case. (b) Schematic of the quasi-isotropic thermal diffusion in nanofibers with large diameters. The molecular chains are randomly oriented and entangled with each other. Heat carriers hop equally to every direction and there is also possibility of inter-chain hopping denoted by red arrow. (c) The setup of eletrospinning. Nanofiber was collected on the two suspended membranes (insert of Figure 1c), which act as heater and temperature sensor during thermal conductivity measurement. (d)Three-dimensional structural map of PI.(e) SEM image of PI nanofiber. The scale bar is 10μm. Red circle marks the position of single PI nanofiber. (f) Enlarged SEM image of the PI nanofiber shown in (e). The scale bar is 30 nm.

Thanks to the development of experimental techniques, it is possible to characterize the thermal conductivity of ultra-thin polymer nanofibers. To test the microstructure dependence of thermal conductivity, it is straightforward to look into the diameter dependence of thermal conductivity in nanofibers through spinning or ultra-drawing[7, 8], during which processes the entanglement of chains could be much reduced by adjusting controllable parameters such as the static-electrical field and draw ratio[29, 30]. In this paper, we systematically investigate the microstructure dependence of thermal conductivity in PI nanofibers for different diameters. The diameter of the obtained PI nanofibers ranges from 31nm to 167nm (see Table S1)and the length of the obtained PI naonfibers are illustrated by Table S2. Molecular chains in thin nanofibers tend to align along the fiber axis with less entanglement as Figure1a demonstrates, while chains in thicker nanofibers are randomly oriented and entangled with each other as is illustrated by Figure 1b.

Figure1c presents the schematic diagram of the electrospinning setup. Due to the static-electrical force introduced by high electrical voltage, the suspended PI nanofibers were formed across the two $SiN_x$ membranes. These two $SiN_x$ membranes were covered by platinum(i.e. the electrical ground). This was the key step where molecular chains tend to align along the axis of the nanofiber. There might be several PI nanofibers passing through the gap of the middle of device after electrospinning. In our experiments, only one nanofiber is left and others will be cut by a nanomanipulator with a tungsten needle(see Figure S1). Figure1d depicts the three-dimensional structural map of PI. It shows large conjugated aromatic bond in a PI structure which could help to enhance the thermal conductivity of PI nanofibers[31].

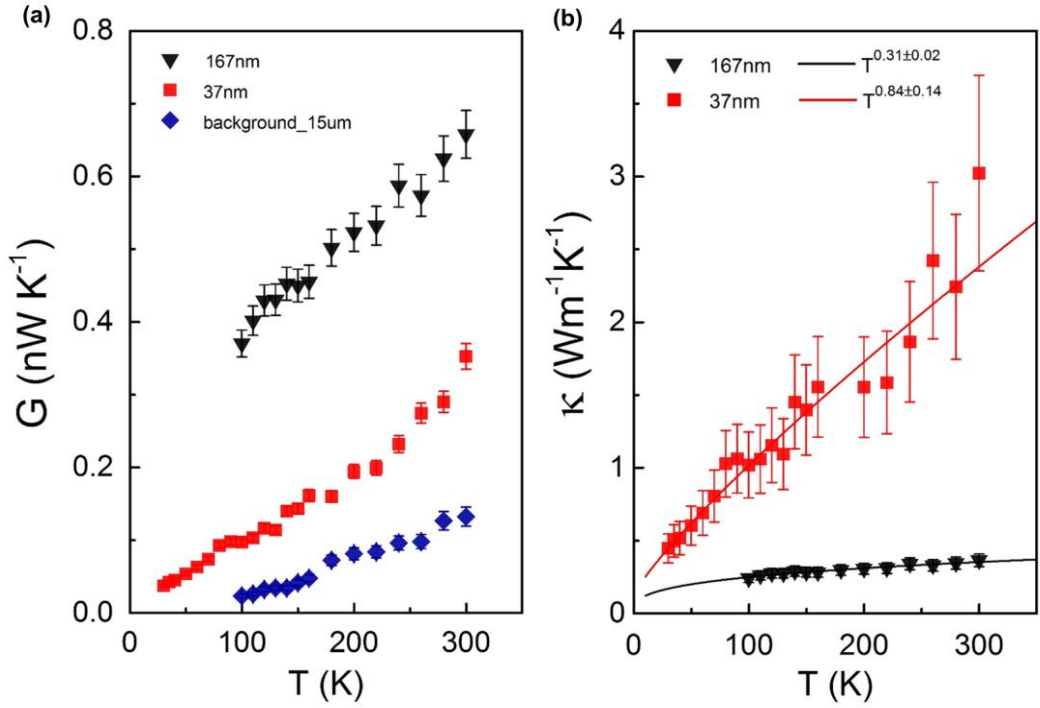

**Figure 2.** Thermal transport of PI nanofibers with different diameters as a function of temperature. (a) Thermal conductance of PI nanofibers. The olive green rhombus points exhibit the thermal radiation measured by the differential circuit configuration with high vacuum (on the order of $1\times10^{-8}$ mbar). (b) Thermal conductivity of PI nanofibers excluding thermal radiation for two samples: No. II $d$=37nm, $L$=14.8um; No. IX $d$=167nm, $L$=15.2um, respectively (the morphology details of other samples were illustrated in Table S1 and S2). Solid lines are fitted by $\kappa \sim T^\lambda$ with $\lambda$=0.84±0.14 and 0.31±0.02 for samples with diameters $d$=37nm and 167nm, respectively. Error bars are estimated based on uncertainties associated with the fiber diameter and temperature uncertainty (see Section S1; Table S1 and S3).

Thermal conductivity along the fiber axis was measured by traditional thermal bridge method[32-34] (Figure 1e and 1f). The whole device was placed in a cryostat with high vacuum on the order of $1\times10^{-8}$ mbar to reduce the thermal convection. To increase the measurement sensitivity, the differential circuit configuration (see Section S2; Figure S2a) was used and the measurement sensitivity of thermal conductance would decrease from ~1nW/K to 10pW/K (see Section S2; Figure S2b)[33, 35]. In our experiment, thermal conductance of PI nanofibers with different diameters is on the order of $1\times10^{-10}$ W/K (Figure 2a). To eliminate the effect of thermal radiation, a blank suspended device was used to probe standard thermal radiation in a wide temperature range. The measured thermal radiation between the two suspended membranes in the blank device is around ~100 pW/K at room temperature. This result is a few times lower than that observed by Pettes *et al.*[36], probably due to better vacuum level which would reduce air conduction and convention. In order to illustrate the effects of thermal contact resistance, two approaches were used to simulate the temperature distribution of the suspended membranes and calculate the thermal contact resistance at the Platinum/PI nanofiber interface. These two approaches verified that the thermal contact resistance held a negligible contribution of the total measured thermal resistance(see Section S3,S4; Figure S3 and S4;Table S4).

The measured thermal conductivity increases monotonously with temperature $T$, which is a typical feature of amorphous material as Figure 2b shows. The amorphous character of PI may arise from the defects and random bond angles within the chain, as well as the complex entanglement between chains. Furthermore, we find that the thermal conductivity varies with temperature as $\kappa \sim T^\lambda$, where $\lambda$ varies from 0.31±0.02 to 0.84±0.14 as the diameter changes. For the thick nanofiber with diameter $d$=167nm, the power law dependence $T^{0.31}$ agrees with the experimental measurement from Singh's group[4]. As the diameter decreases, the power index is approaching to 1, which coincides with the prediction of hopping mechanism[24-26].

To look inside into the intrinsic dominant mechanism of thermal transport, the diameter dependence of thermal conductivity at room temperature is systematically investigated and the results are shown in Figure 3. The thermal conductivity of PI nanofibers is close to that of bulk PI when diameters are larger than 150nm. It increases dramatically as the diameter decreases, and reaches an order of magnitude larger than that in bulk PI when diameters are smaller than 40nm. Similar result was also observed in electrospun Nylon-11 nanofibers[12], which suggested that stretching process could induce more ordered molecular chains in polymers, confirmed by high-resolution wide-angle X-ray scattering.

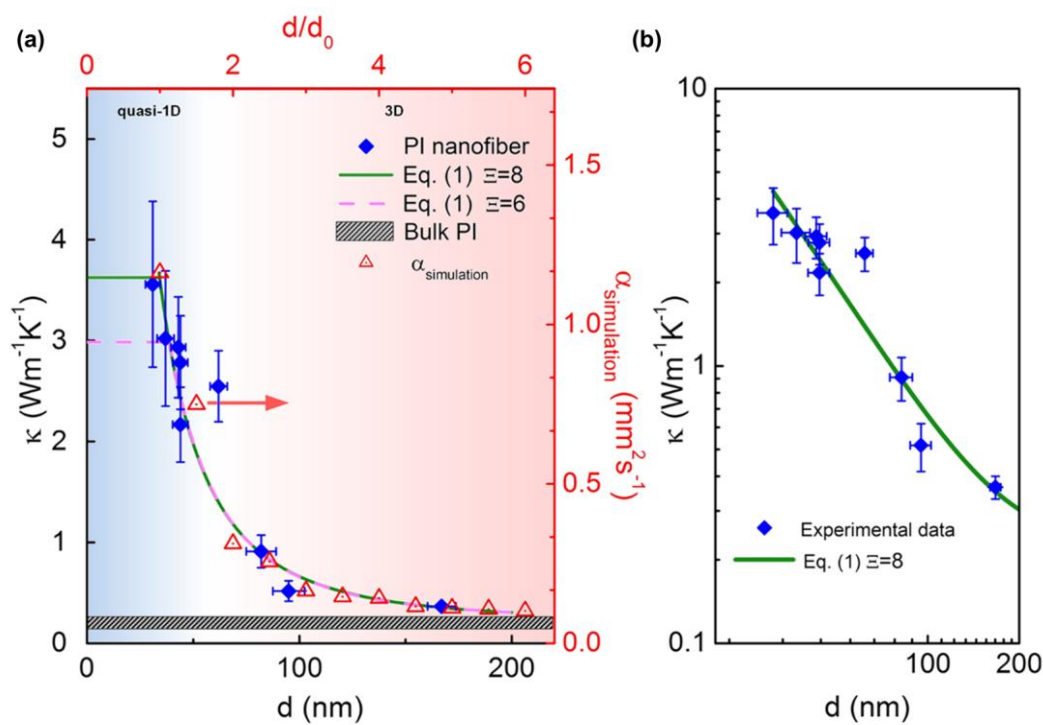

**Figure 3.** Dimensional crossover of thermal conductivity of PI nanofibers at room temperature. (a)The diameter and length detailsof PI nanofibers were illustrated in TableS1 and S2. The gray shadowed bar represents the thermal conductivity of bulk PI within the range of 0.1-0.3Wm$^{-1}$K$^{-1}$. Rhombus (left axis and bottom axis) represent experimental data. Olive solid line (left axis and bottom axis) and pink dash line (left axis and bottom axis) are fitted by equation (1) with different parameter$\Xi$. Red triangles (right axis and upper axis)denote thermal diffusivity obtained from random walk simulation (Details about the random walk simulation are included inFigureS5;Table S5; SectionS5). (b) Dual-logarithm thermal conductivity versus diameter. The olive line is fitted by equation(1).

To describe the diameter dependence of thermal conductivity quantitatively, we propose a theoretical model based on random walk theory to incorporate the diffusion of phonons through hopping mechanism. We are aware of that the lattice vibrations in disordered systems without periodicity do not have dispersion but the terminology of "phonon" is still usable to describe energy quanta. Considering a complex network with a large number of entangled macromolecular chains, phonons transport across this complex network through hopping process. Note that there are two different type of hopping: intra-chain hopping that phonon hops between localization centers within a single chain, as shown in Figure 1a; and inter-chain hopping that phonon hops from one chain to another chain, as shown in Figure 1b. According to the random walk theory, the thermal diffusivity along fiber axis is defined as[37] $\alpha = \frac{1}{\bar{Z}}\Gamma_{tot}\bar{R}^2$, where $\bar{Z}$ is the average effective coordination number along fiber axis, $\bar{R}$ is the average hopping distance and $\Gamma_{tot}$ is the temperature dependent total hopping rate. In our simplified model, we do not consider the difference of hopping rate between inter- and intra-chain hopping processes. Note that the inter-chain hopping process is usually happened at cross links of chains, in which case the hopping distance is negligible compared to that of intra-chain hopping process, it is reasonable to assume that $\bar{R}$ is mainly determined by the intra-chain hopping distance $R_{intra}$. The diameter dependence of $\bar{Z}$ is described by an empirical function $\bar{Z} = [2f(d) + 1][\Xi f(d) + 2]$, where $f(d) = 1 - 2/\{1 + \exp[(d - d_0)/\Lambda]\}$ for $d > d_0$. In the above expression, $d_0$ is the critical diameter under which the diffusion converges to quasi-1D, $\Lambda$ is changing rate of the transition from quasi-1D to 3D and $\Xi$ denotes the average number of inter-chain hopping sites. The average nearest inter-chain neighbor $\Xi$ should be determined from the real configuration of polymer chains. From complex network theory, the number of nearest inter-chain neighbors should be 6~10[38]. For further validation, numerical simulations in generating polymer chains are required. The current form of $f(d)$ could successfully describe the transition from 3D to quasi-1D. When $d = d_0$, $f(d) = 0$ meaning that the system is quasi-1D and there is no inter-chain hopping. When d approaches infinite, $f(d)$ saturates to 1, corresponding to the 3D system. We should stress that our empirical function is definitely not the unique one but it is one of the best ones (like always the case in inverse problem) that fits the experimental data optimally. The thermal conductivity is expressed by $\kappa = \alpha \rho C_p$ where $\alpha$ is thermal diffusivity, $\rho$ is mass density and $C_p$ is specific heat capacity[39], thus thermal conductivity is inversely proportional to $\bar{Z}$ (Details are given in Section S6),

$$\kappa(d) = \frac{2\kappa_{\text{quasi}-1D}}{\bar{Z}}. \tag{1}$$

Note that $f(d) = 0$ when $d \leq d_0$, the thermal conductivity converges to $\kappa_{\text{quasi}-1D}$. It means that the thermal conductivity could not be infinitely increasing with the decreasing of fiber diameter. There exists an upper limit for thermal conductivity of electrospun PI, corresponding to the 1D intra-chain diffusion, where the average effective coordination number along fiber axis is 2. In this limit, all polymer chains are well aligned and the inter-chain interactions are negligible. To verify the validity of our model, we fit the experimental results with equation (1). The fitting parameters are listed in Table 1. Our model fits well with the experimental data as the lines in Figure 3a show.

We also do a random walk simulation and obtain the dimensionless thermal diffusivity. The details of simulation are included in Section S5. The exact value of thermal diffusivity of bulk PI is $\alpha_{bulk} = 0.0775 \text{ mm}^2\text{s}^{-1}$, estimated from the observed thermal conductivity of bulk PI $\kappa_{\text{PI}} = 0.12 \text{ Wm}^{-1}\text{K}^{-1}$, density of bulk PI $\rho_{\text{PI}} = 1.42 \times 10^3 \text{ kg m}^{-3}$, and specific heat capacity of bulk PI $C_{p\,\text{PI}} = 1.09 \times 10^3 \text{ J kg}^{-1}\text{K}^{-1}$ [40]. Nanofibers with diameter smaller than $d_0$ exhibit quasi-1D thermal transport behavior, while nanofibers with diameter much larger than $d_0+\Lambda$ tend to behave like bulk polymers. A crossover of heat conduction from quasi-1D to 3D is only apparent in the range $d_0 \leq d \lesssim d_0 + \Lambda$. The magnitude of critical diameter $d_0$ and parameter $\Lambda$ is related to the radius of gyration $R_g$ of macromolecular chains, which is typically on the order of tenth of nanometers[41]. $R_g$ is determined by the structure of monomers, the bond angle between monomers, the length of single chain, and the process condition such as applied voltage in electronspinning. When $d > d_0 + \Lambda \geq 2R_g$, bulk like polymer nanofibers can be realized since macromolecular chains can easily gyrate. When $d<d_0$, macromolecular chains can hardly gyrate and the chains prefer to lie along the fiber axis.

**Table 1.** Fitting parameters obtained by fitting the experimental data in Figure 3a with equation (1)

| $\Xi$ | $\kappa_{quasi-1D}$ (Wm$^{-1}$K$^{-1}$) | $d_0$ (nm) | $\Lambda$ (nm) |
|---|---|---|---|
| 6 | $3.0 \pm 1.6$ | $39 \pm 13$ | $62 \pm 31$ |
| 8 | $3.6 \pm 2.2$ | $34 \pm 12$ | $66 \pm 36$ |

A crossover of heat conduction from 3D to quasi-1D has been observed experimentally in amorphous polymer nanofibers obtained from electrospinning. This behavior has been quantitatively explained by a model based on random walk theory in which both inter-chain and intra-chain hopping processes are considered. Two important fitting parameters, i.e. $d_0$ and $\Lambda$, are obtained as the characterization length of the dimensional transition. Our theory successfully testifies that the hopping mechanism based on random walk picture is valid and it is useful to explain the diameter dependence of thermal conductivity in nanofibers. Nevertheless, there are still many open questions deserving requiring further investigations. First, the temperature dependence of thermal conductivity has not been well explained and it requires deeper and quantitative understanding of the inter-chain thermal transport mechanism. Second, the four parameters in the empirical function require validation from further simulations and experiments. For example, $\kappa_{\text{quasi}-1\text{D}}$ is closely related to the thermal conductivity of single chain, and it is possible to be obtained from molecular dynamics; $\Xi$, $d_0$ and $\Lambda$ are determined by the configuration of polymer chains which requires research on polymer condensed matter physics and thermal measurements on much thinner polymer fibers.

**Experimental Section**

**Thermal conductivity measurement**

The PI nanofibers fabricated by electrospinning method were served as bridge to connect two Platinum/SiN$_\text{x}$ membranes (Figure1e). These two membranes were regarded as thermometers. A DC current of a slow change step combined with an AC current (1000nA) was added to one of the membranes served as heater resistor ($R_\text{h}$, the left Platinum coil shown in Figure 1c). The DC current was applied to provide Joule heat and also to increase its temperature ($T_\text{h}$). The AC current was used to measure the resistance of $R_\text{h}$. Meanwhile, an AC current of the same value was applied to another membrane served as sensor resistor ($R_\text{s}$, the right Platinum coil shown in Figure 1c), to probe the resistance of $R_\text{s}$. The Joule heating in $R_\text{h}$ gradually dissipated through the six Platinum/SiN$_\text{x}$ beams and the PI nanofiber, which would raise the temperature in $R_\text{s}$ ($T_\text{s}$). At steady state, the thermal conductance of PI nanofibers ($\sigma_\text{PI}$) and the thermal conductance of suspended beam ($\sigma_\text{l}$) could be obtained by

$$\sigma_\text{l} = \frac{Q}{\Delta T_\text{h} + \Delta T_\text{s}}$$

and

$$\sigma_\text{PI} = \frac{\sigma_\text{l}\Delta T_\text{s}}{\Delta T_\text{h} - \Delta T_\text{s}}$$

where, $\Delta T_h$ and $\Delta T_s$ indicate the temperature rise in the $R_h$ and $R_s$, Q is the Joule heat applied to the heater resistor and one of the Platinum/SiNx beams.

**Electrospinning**

To fabricate nanoscale PI fibers with controllable diameter and chain orientation, we utilized eletrospinning using a commercialized electrospinner. The solvent, mixture of PI and Dimethylformamide (DMF) solution, was prepared with concentration from 45% to 80%, followed by all night stirring to guarantee PI molecules and DMF solvent mixed completely. The diameter of PI nanofiber should increase with the increasing of PI molecules weight ratio.

**ASSOCIATED CONTENT**

**Supporting Information**

Measurement error bar, the differential circuit configuration, Finite Element Simulations (COMSOL Multiphysics 5.2), thermal contact resistance, Random walk simulation, diameter dependence of average coordination number,

**AUTHOR INFORMATION**

**Corresponding Author**

*E-mail: zhoujunzhou@tongji.edu.cn

*E-mail: xuxiangfan@tongji.edu.cn

*E-mail: Baowen.Li@Colorado.edu

**Notes**

The authors declare no competing financial interests.

ACKNOWLEDGMENT

The work was supported by the National Natural Science Foundation of China (No. 11674245 & No. 11334007 & No. 11505128), by the program for Professor of Special Appointment (Eastern Scholar) at Shanghai Institutions of Higher Learning (No. TP2014012), and by Shanghai Committee of Science and Technology in China (No. 17142202100 & No. 17ZR1447900). Special thanks to the technical engineer Ning Liu from ZEISS who helped to obtain the high resolution SEM image.

# Supporting Information
# for
## "Dimensional crossover of heat conduction in Amorphous Polyimide Nanofibers"

*Lan Dong[1,2,3†], Qing Xi[1,2,3†], Dongsheng Chen[4,5], Jie Guo[1,2,3], Tsuneyoshi Nakayama[1,2,3,6], Yunyun Li[1,2,3], Ziqi Liang[4], Jun Zhou[1,2,3\*], Xiangfan Xu[1,2,3\*], Baowen Li[7\*]*

[1]*Center for Phononics and Thermal Energy Science, School of Physics Science and Engineering, Tongji University, 200092 Shanghai, China*

[2] *China-EU Joint Center for Nanophononics, School of Physics Science and Engineering, Tongji University, 200092 Shanghai, China*

[3]*Shanghai Key Laboratory of Special Artificial Microstructure Materials and Technology, School of Physics Science and Engineering, Tongji University, 200092 Shanghai, China*

[4]*Department of Materials Science, Fudan University, 200433 Shanghai, China*

[5]*College of Mathematics and Physics, Shanghai University of Electric Power, 200090 Shanghai, China*

[6]*Hokkaido University, Sapporo 060-0826, Japan*

[7]*Department of Mechanical Engineering, University of Colorado, Boulder, CO 80309-0427, USA*

[†]These authors contributed equally to this work

**Table of contents:**

**Section S1.Thermal conductivity uncertainty**

The error bars are determined by:

1. Temperature uncertainty (~5%)

    We used the same experimental setup as P.Kim and L.Shi[1], in which the thermal conductivity of single nanotube wasmeasured for the first time. This measurement setup has been tested to be one of the most successful techniques in measuring thermal transport properties of single nanotube, nanowire and nanofiber. In our experiments, the temperature uncertainty is between 1.8% to 2.7% in all samples(see Table S3). Considering the effects of system noise and temperature fluctuation of the base temperature in cryostat, we simply assume 5% temperature uncertainty in our experiments.

2. Error bars of thermal conductivity

    The thermal conductivity uncertainty of PI nanofibers is composed of the temperature uncertainty and the diameter uncertainty of PI nanofibers. Here, the diameter uncertainty is measured by SEM(see Table S1). The uncertainty in the thermal conductivity of each PI nanofiber is estimated using formula as follow:

    1. $$\frac{\delta \kappa}{\kappa} = \sqrt{(\frac{\delta G}{G})^2 + (2\frac{\delta d}{d})^2}$$

    Where $\delta \kappa$ is the uncertainty in the thermal conductivity, $G$ is the thermal conductance of PI nanofibers, $\delta G$ is the uncertainty in the thermal conductance which is induced by temperature uncertainty in our experiment. $d$ is the diameter of PI nanofibers, $\delta d$ is the uncertainty in the

diameter of PI nanofibers when measured by SEM. Therefore, the uncertainty in thermal conductivity is from 9% to 23% in all samples. The high uncertainty in thermal conductivity up to 23% is due to the increased difficulty in measuring the diameters under SEM when the diameter of the nanofiber is less than 40nm. It should be emphasized that the uncertainty in all sample are within expectations.

**Section S2.The differential circuit configuration**

Our experiments adopt a differential circuit configuration(Figure S2a) to offset the nominal resistance of the sensor membrane which could help to achieve a higher resolution. The measurement results of the differential circuit configuration in our experiments are shown in Figure S2b. Black dashed line represents the results of standard thermal bridge method whose resolution fluctuation approach to 100mK(this result is directly relevant to the temperature drift, the stability of power supply as well as the grounding of system.); Red dot line represents the data of the differential circuit configuration and its accuracy is about 5mK. Simultaneously we found the temperature drift of cryostat is about 70mK in 10 minutes. It has been clarified that the temperature drift of the system could affect the experimental error, especially for the high resolution measurement. To this end, we place a variable resistor ($R_f$) in series with the sensor in the cryostat. In this case, the temperature drift could affect both the sensor membrane and the variable resistor ($R_f$). When the sensor resistance is consistent with resistor resistance($R_f$), the effect of temperature drift could be negligible, as shown in the Figure S2b, the blue solid line represents the measurement resolution approach to 2mK. Because of this high measurement resolution, the measured sensitivity of thermal conductance decreases to 10pW/K.

**Section S3. Finite Element Simulations (COMSOL Multiphysics 5.2)**

The Figure S3 shows temperature distribution in the suspended structure ($R_h$, $R_s$ and the PI nanofiber). Figure S3a exhibits the schematic of the structure. The temperature rise on the heater membrane is 9.797K. Here, the temperature distribution changes within 0.2% both on $R_h$ and $R_s$ membranes (shown in the Figure S3b), thus we could take the temperature distribution as a constant and need no more correction on the measured temperature rise ($\Delta T_h$, $\Delta T_s$) in our experiments. Figure S3c and S3d illustrate the temperature distribution on the suspended PI nanofiber.

In order to illuminate the effects of thermal contact resistance in our experiments, the Finite Element Simulation was carried out to calculate thermal resistance of the suspended PI nanofiber. The simulation results were exhibited in Figure S4. The thermal conductivity of platinum could be obtained from the Weidemann-Franz law and its thermal conductivity is calculated to be 23.2 Wm$^{-1}$K$^{-1}$, based on following simulation.

For the simulation, the temperature distribution of the suspended structure is calculated for different set values $\kappa_s$ (the set values of thermal conductivity in PI nanofiber ranging from 0.02Wm$^{-1}$K$^{-1}$ to 100Wm$^{-1}$K$^{-1}$). The thermal resistance of PI nanofiber(d=37nm) $R_m$(=1/$G_m$) is plotted as function of the set value $\kappa_s$, from which the value $G_m$ can be obtained as $G_m = Q \times \Delta T_s / (\Delta T_h^2 - \Delta T_s^2)$, where Q is the Joule heat applied to the heater, $\Delta T_h$ and $\Delta T_s$ are average temperature rise on the two membranes which was obtained from simulation.

Here we can see, the $R_m$ contains both sample resistance and total thermal contact resistance, and $R_s$ could be calculated from $\kappa_s$. When $R_s$ close to zero, the y-axis intercept could represent the total thermal contact resistance. In Figure S4, $R_m$ shows a linear dependence on $R_s$. The total thermal contact resistance approach to $1.31 \times 10^7$ K/W at 300K and increase to $1.28 \times 10^8$ K/W at 50K. Based on our model, the thermal contact resistance contribute <1% to the total measured thermal resistance of sample in a wide temperature range.

**Section S4. Thermal contact resistance**

To be consistent and eliminate the sample-to-sample variation in contact resistance, the data were collected from the same batch of Polyimide solution and also from the same $SiN_x$ wafers, each of them integrated with 6×7 devices array with different gaps.

The thermal contact resistance $R_c$ between the PI nanofibers and the two membranes ($R_h$ and $R_s$) can be calculated using a fin thermal resistance model[2,3]:

$$R_c/2 = \left[\sqrt{\frac{\kappa A w}{R_{interface}}} \tanh\left(\sqrt{\frac{w}{\kappa A R_{interface}}} l_c\right)\right]^{-1}$$

where $R_c$ is thermal contact resistance, the number 2 in $R_c/2$ indicate two contacts on the two sides of PI nanofibers, $\kappa$ is thermal conductivity of PI nanofibers, A is the contact area between sample and electrode, w is the sample width, $l_c$ is the contact length, and $R_{interface}$ is the thermal interfacial resistance.

Here, we assume that room temperature thermal interfacial resistance of PI/platinum/$SiN_x$ interface in our devices is within the range from $6.4 \times 10^{-8}$ m$^2$KW$^{-1}$ to $9.7 \times 10^{-8}$ m$^2$KW$^{-1}$ in metal-organic interface[4]. From this, the thermal contact resistance($R_c$) in our samples is calculated to be ~$1.9 \times 10^7$ KW$^{-1}$ to $1.7 \times 10^7$ KW$^{-1}$ which accounts for ~0.6% to 2.7% of the total measured thermal

resistance($R_{tot}$) in samples at room temperature. The contribution of the thermal contact resistance to the total thermal resistance of each sample been shown in Table S4.

**Section S5. Random walk simulation**

A random walk simulation has also been done to check the validity of our model. Instead of building a complex structure of entangled chains, we manipulate it to the inter-chain coordination number $\bar{Z}_{inter}(\tilde{d}) = \Xi f(\tilde{d})$ and confinement effect in radius direction. The confinement in radius direction is characterized by the reduced diameter $\tilde{d} = d/d_0$ and it is realized by an elastic collision boundary condition. Every step we generate a random number to determine the next position, and after N steps, we calculate the squared displacement $x^2$ along the fiber axis from the start point. Another 10000 times of the previous procedure are done to obtain the average $\langle x^2 \rangle$, and thermal diffusivity is calculated by $\alpha = \frac{\langle x^2 \rangle}{2N}$. The diameter confinement effect could be explained by Figure S5 and Table S5. The chain segment could hardly oriented perpendicular to the fiber axis because of the confinement of boundary when $\tilde{d} \to 1$, so the random walk is always along the fiber axis, which leads to a larger thermal diffusivity along the fiber axis. With the increasing of $\tilde{d}$, the confinement becomes weaker and segments oriented isotropically in the fiber, which leads to a smaller thermal diffusivity along fiber axis.

**Section S6. Diameter dependence of average coordination number**

Let's consider a phonon at a certain position ($r_i$) of a chain noted as A, the local coordination number is defined as $Z(r_i) = Z_{inter}(r_i) + Z_{intra}(r_i)$, which is the sum of the inter-chain component $Z_{inter}(r_i)$ and the intra-chain component $Z_{intra}(r_i)$. There are always two possible hopping events along chain A, i.e. forward and backward, which means $Z_{intra}(r_i) \equiv 2$. If there is no other chain entangled with chain A near $r_i$ (from $r_i - R_{intra}/2$ to $r_i + R_{intra}/2$), $Z(r_i) = Z_{intra}(r_i) = 2$. If there is another chain (B) entangled with chain A near $r_i$, then $Z(r_i) = 3$ since $Z_{inter}(r_i) = 1$ which comes from the possible inter-chain hopping at the cross link. Therefore, $\bar{Z} = \sum_i Z(r_i) / \sum_i p_i$, where $Z(r_i) = 2 + Z_{inter}(r_i)$ and $p_i$ is the probability that the segment from $r_i - R_{intra}/2$ to $r_i + R_{intra}/2$ is along the fiber axis. For ultra-thin nanofibers, most macromolecular chains tend to align along the nanofiber as shown in Figure 1a that $p_i \to 1$. Also, very few entanglements lead to very few cross links that $Z_{inter}(r_i) \to 0$. The thermal diffusivity will converge to the quasi-1D random walk results, $\alpha_{quasi-1D} = \frac{1}{2}\Gamma_{tot}\bar{R}^2$. For isotropic bulk polymer and thick nanofibers as shown in Figure 1b, macromolecular chains are randomly oriented that $p_i = 1/3$. Also, the chains are entangled with each other that each segment may have $\Xi$ cross links, $Z_{inter}(r_i) \to \Xi$. The thermal diffusivity will converge to the 3D isotropic random walk results, $\alpha_{iso} = \frac{1}{3(\Xi+2)}\Gamma_{tot}\bar{R}^2$. $\Xi$ is a fitting parameter because the detailed topological structure of the localized center is unclear. $\Xi = 0$ results in $\bar{Z} = 6$ for a simple isotropic system without close entanglement. It is obvious that the thermal diffusivity is determined by the average coordination number, which is related to 2 components: the dimensionality and the structure of the network, both of which are sensitively dependent on the diameter of the nanofibers. Here, we introduce an empirical function $f(d)$ to characterize the diameter confinement effect. $f(d)$ will change from 0 to 1 when d increases to infinity. As a result, the diameter dependence of $p_i$ can be described by $\bar{p}_i = \frac{1}{2f(d)+1}$, while the diameter dependence of $Z(r_i)$ is $\overline{Z(r_i)} = \Xi f(d) + 2$. The average coordination number is thus expressed by $\bar{Z} = (2f(d) + 1)(\Xi f(d) + 2)$.

Supposing that the only difference between nanofibers with different diameters lies in the average coordination number $\bar{Z}$, the thermal conductivity of nanofibers with arbitrary diameter could be determined from $\frac{\kappa_{quasi-1D}}{\kappa(d)} = \frac{2}{\bar{Z}_{quasi-1D}}$, which leads to equation (1) in the main text.

Considering the fact that the interactions between chains are Van der Waals forces or other weak interactions, which are weaker than intra-chain covalent bonds, it is oversimplified to treat equally the intra-chain diffusion and inter-chain diffusion. For those who want to treat accurately the difference between intra-chain diffusion and inter-chain diffusion, a more detailed derivation of $\bar{Z}$ is given following

Since we are now discussing uncorrelated random walk that each individual walk is independent of all prior walks, the equation describing a random walk without correlation is given by[5]

$$\langle R^2 \rangle = \sum_{i=1}^{n} \langle r_i^2 \rangle$$

$$\langle X_{fiber}^2 \rangle = \sum_{i=1}^{n} \langle x_{fiber,i}^2 \rangle$$

where $R$ is the total displacement and $r_i$ is individual displacement of each step. $X_{fiber}$ is the total displacement along the fiber axis, and $x_{fiber,i}$ is the displacement along fiber axis of each individual step.

In a given configuration of polymer chains, the heat carriers could either jump to an intra-chain site (jump length $r_{intra}$ with component $x_{fiber,intra}$ along the fiber axis) or jump to another chain (jump length $r_{inter}$ with component $x_{fiber,inter}$ along the fiber axis). Therefore, the above two equations reduce to

$$\langle R^2 \rangle = \langle n_{intra} \rangle r_{intra}^2 + \langle n_{inter} \rangle r_{inter}^2$$

$$\langle X_{\text{fiber}}{}^2 \rangle = \langle n_{\text{intra}} \rangle x_{\text{fiber,intra}}^2 + \langle n_{\text{inter}} \rangle x_{\text{fiber,inter}}^2$$

here $\langle n_{\text{intra}} \rangle$ and $\langle n_{\text{inter}} \rangle$ denote the average number of jumps of one heat carrier inside the chain or between chains, respectively. The total jumping rate is given by

$$\Gamma_{\text{total}} = \frac{\langle n_{\text{intra}} \rangle + \langle n_{\text{inter}} \rangle}{T}$$

where T is the total time. The thermal diffusivity along fiber axis is defined as

$$\alpha = \frac{\langle X_{\text{fiber}}{}^2 \rangle}{2T} = \frac{\langle n_{\text{intra}} \rangle}{2T} x_{\text{fiber,intra}}^2 + \frac{\langle n_{\text{inter}} \rangle}{2T} x_{\text{fiber,inter}}^2$$

Effective coordination numbers are introduced here to express the thermal diffusivity by the total jumping rate,

$$\alpha = \left( \frac{1}{\bar{Z}_{\text{intra}}^{\text{eff}}} \bar{x}_{\text{fiber,intra}}^2 + \frac{1}{\bar{Z}_{\text{inter}}^{\text{eff}}} \bar{x}_{\text{fiber,inter}}^2 \right) \Gamma_{\text{total}}$$

where

$$\bar{Z}_{\text{intra}}^{\text{eff}} = 2\left(1 + \frac{\langle n_{\text{inter}} \rangle}{\langle n_{\text{intra}} \rangle}\right)$$

$$\bar{Z}_{\text{inter}}^{\text{eff}} = 2\left(1 + \frac{\langle n_{\text{intra}} \rangle}{\langle n_{\text{inter}} \rangle}\right)$$

and the difference between intra-chain interaction and inter-chain inter action could be included in the effective coordination number by term $\frac{\langle n_{\text{inter}} \rangle}{\langle n_{\text{intra}} \rangle}$.

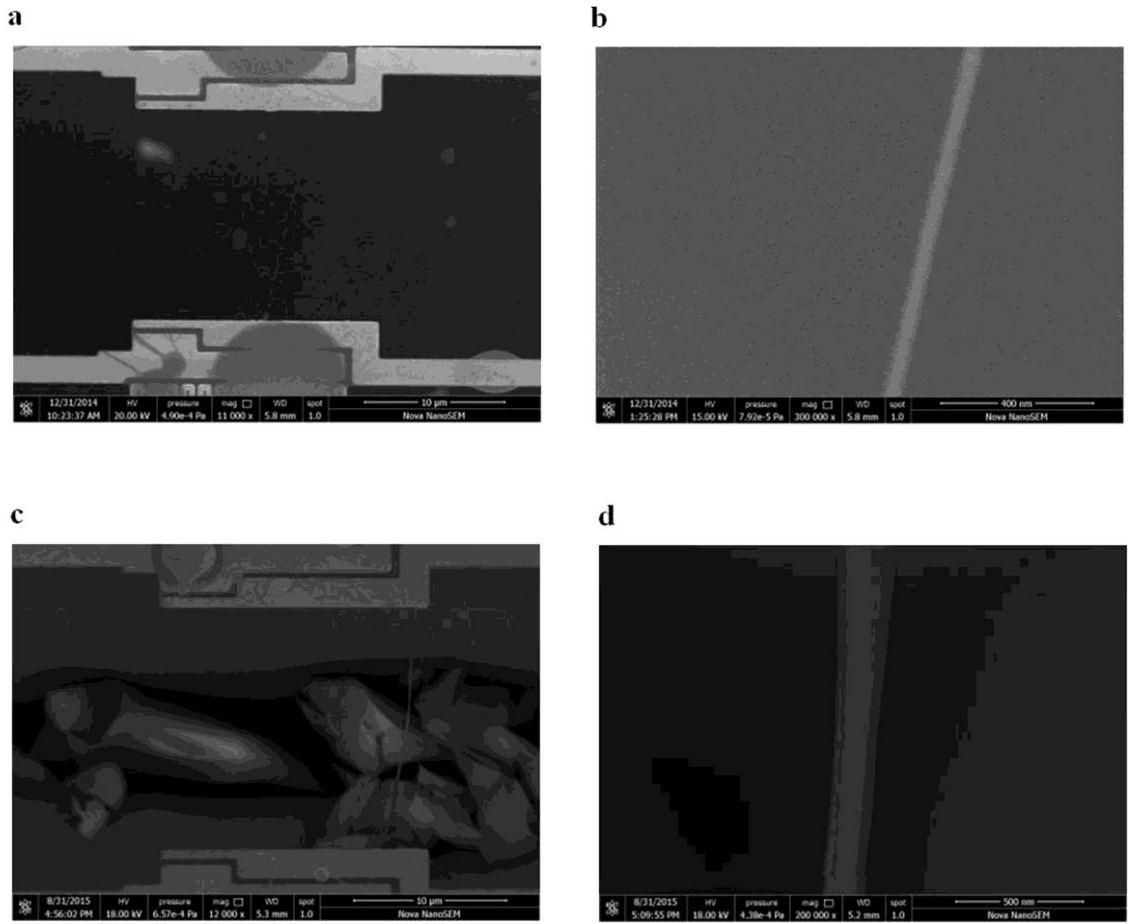

**Figure S1.** Sample details for PI nanofibers. (a) A single PI nanofibers passing through the gap in the middle of suspended device with diameter in 37nm, and length of this nanofiber is 14.8um. The scale bar is 10um. (b) Enlarged SEM image of the same PI nanofiber shown in (a), the scale bar is 400nm. (c) A single PI nanofibers with diameter in 167nm. and length of this nanofiber is 15.2um. The scale bar is 10um. (d) Enlarged SEM image of the same PI nanofiber shown in (c), the scale bar is 500nm.

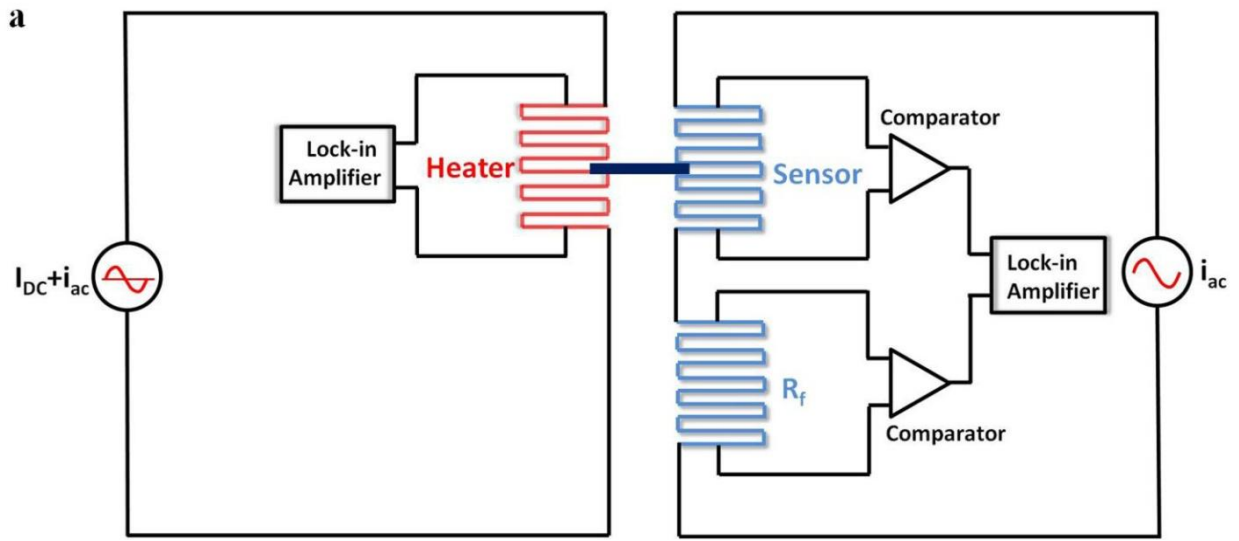

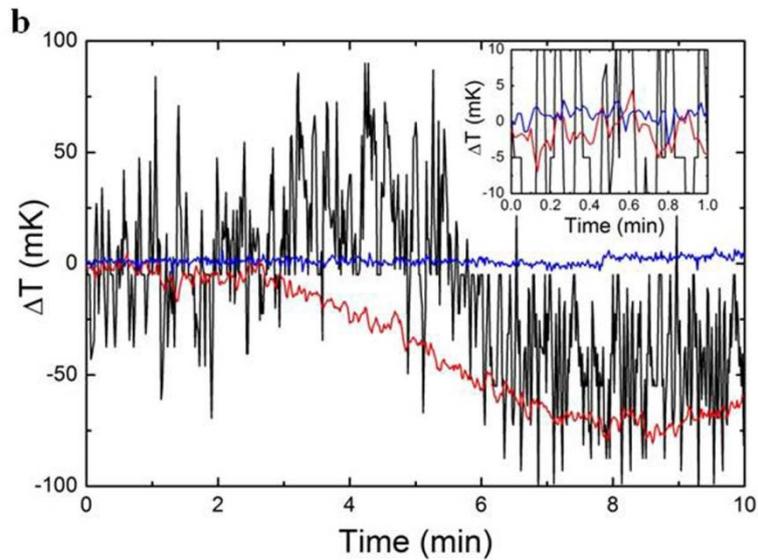

**Figure S2.** The differential circuit configuration and temperature resolution measured. (a) The differential circuit configuration is used to offset the nominal resistance of the Platinum/SiNx membrane($R_s$). (b) The temperature resolution measured by differential circuit configuration.

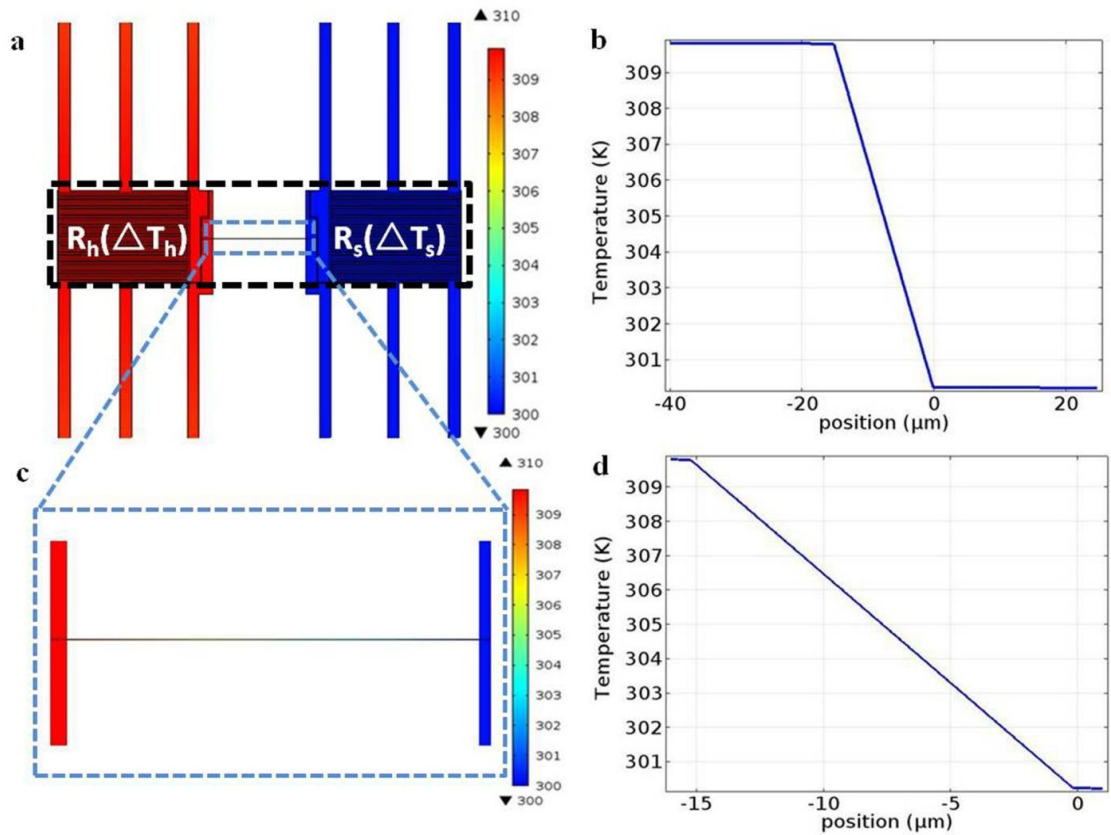

**FigureS3.** Finite Element Simulations (COMSOL Multiphysics 5.2, License No: 9400382) of the temperature distributions in sample device. (a) Schematic of sample device in finite element simulation. (b) The temperature distribution of the rectangle area with black dashed border. The black dashed rectangle was demonstrated in (a). (c) Enlarged schematic of PI nanofiber which was placed across the middle gap of device (as the blue dashed rectangle demonstrates in (a)). (d)The temperature distribution of the PI nanofiber. The temperature rise on the heater membrane ($R_h$) is 9.797K. The temperature distribution change is within 0.2% both on the heater ($R_h$) and sensor ($R_s$) membranes.

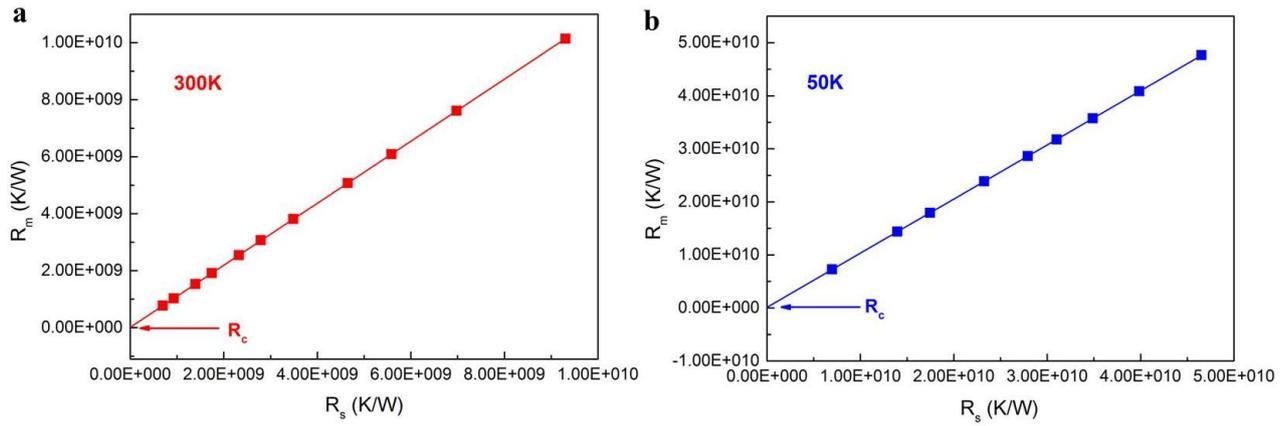

**FigureS4.** Calculation of thermal contact resistance. The calculated thermal resistance $R_m$ of suspended PI nanofiber (d=37nm) with respect to the set value $R_s$ with the best fit line. The figures illustrate that the thermal contact resistance contribute <1% to the total measured thermal resistance of sample in a wide temperature range.

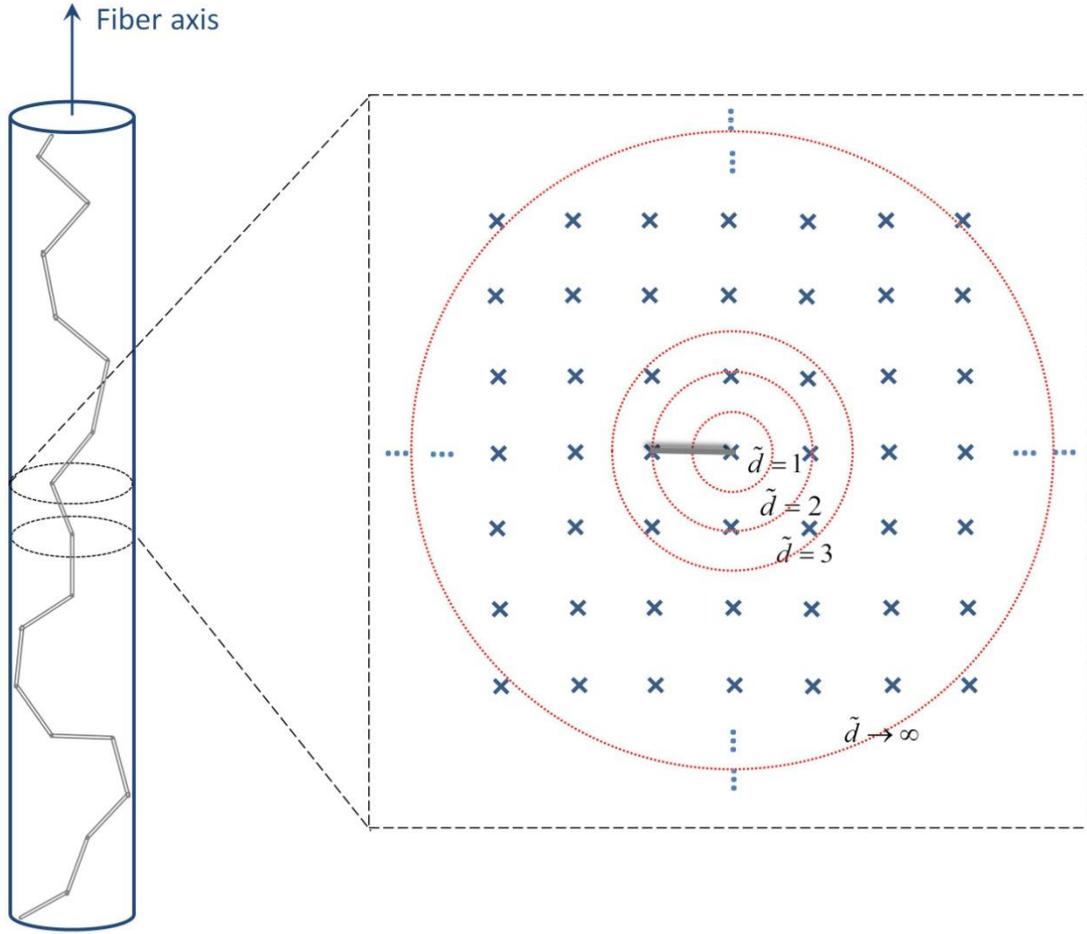

**Figure S5.** Schematic of the diameter confinement effect on the coordination number. Molecular chains are formed by chain segments as the left figure shows. The confinement effect on the conformation of molecular chains is illustrated by the boundary confinement as the right figure shows. When the reduced diameter $\tilde{d} \leq 1$, chain segments could only align along the fiber axis which leads to a coordination number of $Z_{intra} = 2$. With the increasing of reduced diameter, the boundary confinement is weaker and the average effective coordination number will approach to the situation of free jointed chain that $Z_{intra}^{eff} = 6$. The average effective coordination number along fiber axis $\bar{Z}_{intra}^{eff}$ corresponding to different reduced diameter $\tilde{d}$ is listed in Table S5.

**Table S1.** Diameter uncertainty of PI nanofibers.

| No. | I | II | III | IV | V | VI | VII | VIII | IX |
|---|---|---|---|---|---|---|---|---|---|
| Sample diameter(nm) | 31 | 37 | 43 | 44 | 44 | 62 | 82 | 95 | 167 |
| Diameter uncertainty(nm) | 3.5 | 4 | 3.5 | 3.6 | 3.5 | 4 | 7 | 7.5 | 6.7 |
| Uncertainty in κ | 23% | 22% | 17% | 17% | 17% | 14% | 18% | 17% | 9% |

**Table S2.** The length of PI nanofibers fabricated by electrospining.

| No. | I | II | III | IV | V | VI | VII | VIII | IX |
|---|---|---|---|---|---|---|---|---|---|
| Sample diameter(nm) | 31 | 37 | 43 | 44 | 44 | 62 | 82 | 95 | 167 |
| Sample length(um) | 20.0 | 14.8 | 20.0 | 20.2 | 20.2 | 15.0 | 15.0 | 21.2 | 15.2 |

**Table S3.** Measurement uncertainty.

| Sample diameter(nm) | 37 | | 62 | | 167 | |
|---|---|---|---|---|---|---|
| Temperature | T=30K | T=300K | T=30K | T=300K | T=100K | T=300K |
| $\Delta T_S$ | 8mK | 33mK | 65mK | 87mK | 60mK | 71mK |
| Uncertainty | 2.7% | 1.9% | 1.4% | 1.4% | 2.3% | 1.8% |

**Table S4.** The proportion of thermal contact resistance in each sample.

| When concerning $R_{interface}=6.4\times10^{-8} m^2 KW^{-1}$ | | | | | |
|---|---|---|---|---|---|
| Diameter(nm) | 31 | 37 | 43 | 44 | 44 |
| $R_C/R_{tot}$ | 1.1% | 1.6% | 1.1% | 0.9% | 1.0% |
| Diameter(nm) | 62 | 82 | 95 | 167 | |
| $R_C/R_{tot}$ | 2.2% | 1.2% | 0.6% | 1.0% | |

| When concerning $R_{interface}=9.7\times10^{-8} m^2 KW^{-1}$ | | | | | |
|---|---|---|---|---|---|
| Diameter(nm) | 31 | 37 | 43 | 44 | 44 |
| $R_C/R_{tot}$ | 1.3% | 1.9% | 1.4% | 1.2% | 1.4% |
| Diameter(nm) | 62 | 82 | 95 | 167 | |
| $R_C/R_{tot}$ | 2.7% | 1.4% | 0.8% | 1.3% | |

**Table S5.** Effective intra-chain coordination numbers along fiber axis $\bar{Z}_{intra} = \sum_i Z_{intra}(r_i)/\sum_i 1$ versus reduced diameter $\tilde{d}$.

| Reduced diameter $\tilde{d}$ | Average effective intra-chain coordination numbers $\bar{Z}_{intra}^{eff}$ |
|---|---|
| 1 | 2 |
| 2 | 3.6 |
| 3 | 4.5 |
| 4 | 5.0 |
| 5 | 5.3 |
| … | … |
| ∞ | 6 |